\documentclass[final]{vgtc}                 

\ifpdf
  \pdfoutput=1\relax                   
  \usepackage{graphicx}                
  \DeclareGraphicsExtensions{.pdf,.png,.jpg,.jpeg} 
\else
  \usepackage{graphicx}                
  \DeclareGraphicsExtensions{.eps}     
\fi%

\graphicspath{{figures/}{pictures/}{images/}{./}} 

\usepackage{xcolor}
\usepackage{microtype}                 
\PassOptionsToPackage{warn}{textcomp}  
\usepackage{textcomp}                  
\usepackage{mathptmx}                  
\usepackage{times}                     
\usepackage{cite}                      
\usepackage{tabu}                      
\usepackage{booktabs}                  
\usepackage[english]{babel}
\newcommand{\etal} {\textit{et al.}}

\onlineid{1244}

\vgtccategory{Application}

\vgtcinsertpkg

\title{A Research Framework for Virtual Reality Neurosurgery Based on Open-Source Tools}

\author{Lukas D.J. Fiederer\thanks{e-mail: lukas.fiederer@uniklinik-freiburg.de}\\ %
     \parbox{1.7in}{\scriptsize \centering Translational Neurotechnology Lab, Medical Center - University of Freiburg, Germany} %
\and Hisham Alwanni\\
     \parbox{1.7in}{\scriptsize \centering Translational Neurotechnology Lab, Medical Center - University of Freiburg, Germany}%
\and Martin V\"olker\\
     \parbox{1.7in}{\scriptsize \centering Translational Neurotechnology Lab, Medical Center - University of Freiburg, Germany\\~\\}
\and Oliver Schnell\\
    \parbox{1.7in}{\scriptsize \centering Department of Neurosurgery, Medical Center - University of Freiburg, Germany}
\and J\"urgen Beck\\
    \parbox{1.7in}{\scriptsize \centering Department of Neurosurgery, Medical Center - University of Freiburg, Germany}
\and Tonio Ball\\
     \parbox{1.7in}{\scriptsize \centering Translational Neurotechnology Lab, Medical Center - University of Freiburg, Germany
     }}

\abstract{

Fully immersive virtual reality (VR) has the potential to improve neurosurgical planning. For example, it may offer 3D visualizations of relevant anatomical structures with complex shapes, such as blood vessels and tumors. However, there is a lack of research tools specifically tailored for this area. 
We present a research framework for VR neurosurgery based on open-source tools and preliminary evaluation results. We showcase the potential of such a framework using clinical data of two patients and research data of one subject. 
As a first step toward practical evaluations, two certified senior neurosurgeons positively assessed the usefulness of the VR visualizations using head-mounted displays. The methods and findings described in our study thus provide a foundation for research and development aiming at versatile and user-friendly VR tools for improving neurosurgical planning and training.
} 

\CCScatlist{
    Virtual reality, Imaging, Health informatics, Image segmentation, Medical technologies, 3D imaging.
}

\begin{document}



\firstsection{Introduction}

\maketitle

As of 2019, neurosurgeons still plan many interventions using 2D images on 2D screens to build an internal 3D representation of what to expect during surgery and of how to proceed to achieve best possible outcomes. However, this process is time-consuming, requires extensive experience, and mentally reconstructing the relevant complex 3D structures, such as blood vessels and their relation to brain tumors, from the 2D data may lead to inaccuracies. Virtual Reality (VR) offers the possibility for an immersed experience of 3D content, and thus gives neurosurgeons the means of a direct 3D perception of and interaction with the anatomy relevant for planning.  
A crucial step for applying VR to neurosurgery is the precise segmentation needed to create sufficiently accurate models (as required by neurosurgeons) based on imaging data recorded in the clinical routine. Most current pipelines are based on research-grade data (higher signal-to-noise and spatial resolution compared to routine clinical data) and their applicability to clinical routine data is questionable. 

For example, previously, we published 
a high-end segmentation framework for the creation of whole-head sub-mm resolved models~\cite{fiederer2016role}. This framework is based entirely on open-source tools, and has been shown to be especially effective for segmenting the intricate blood vessel trees supplying the brain and the surrounding structures, as well as being able to process imaging data at a finer resolution than the widely used 1-mm isotropic resolution. Knowledge about the exact position of even small blood vessels is crucial for the planning of neurosurgical interventions.

However, this segmentation pipeline was designed and evaluated for research-grade MRI images of healthy individuals and conjoint with conventional 3D visualization. Therefore, it is unclear how this processing approach could be translated to MRI and CT images acquired during clinical routine and as a basis for immersive visualization in VR. As described above, neurosurgeons have to create a complex internal representation of the surgical situation and this process can potentially be eased by VR technology. So far, while first commercial products are currently being developed in this area, an open-source immersive VR framework is not available for this task, nor therefore, for research and development in this application context. However, using open-source tools in research is important to ensure that the work can be reproduced properly and can be used as a foundation for further work by the community.

In relation to this we have started to addressed the following questions and report our preliminary results: 1) Can a versatile and user-friendly VR framework for neurosurgery be implemented using open-source tools? 2) How can the existing segmentation framework be adapted to allow reliable segmentation of cranial structures, including pathological changes, from clinical imaging data obtained in neurosurgical patients, for subsequent visualization in VR? \textcolor{red}{3}) In particular, how can individual differences, such as in the type and quality of the available imaging data, be accounted for in the data processing approach? 

\begin{figure*}[h!]
\centering{\includegraphics[width=\textwidth]{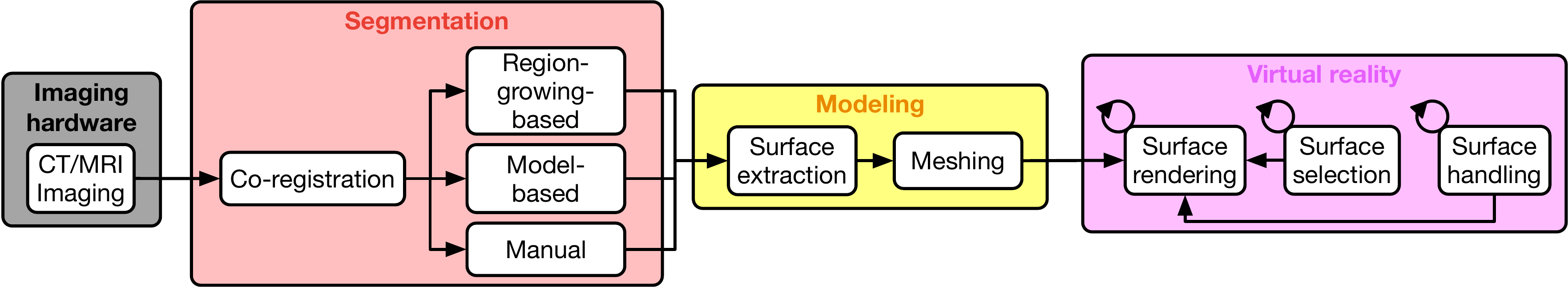}}
\caption[A Virtual Reality Framework for Neurosurgery]
{A Virtual Reality (VR) Framework for Neurosurgery: Current state of our framework for providing realistic 3D models of patient specific data to neurosurgeons in fully immersive VR (visual and auditory senses are fully captured by the VR). The arrows indicate the flow of data/information in the framework. Circling arrows indicate recurrent steps in time. Significant improvements are still planned.
}
\label{fig:Framework}
\end{figure*}

To address these questions, we propose a research framework for VR neurosurgery, based on open-source tools, for the transformation of patient-specific imaging data into VR 3D representations (Fig. \ref{fig:Framework}). Our framework extends the conventional planning of surgical interventions by segmenting and modeling patient-specific anatomical structures in 3D and integrating them into an immersive 3D VR environment. Surgeons can use head-mounted displays (HMDs) and wireless controllers to interact with the modeled imaging data in real-time (smoothly, without any time delays) via a user-friendly interface. The interaction capabilities include, but are not limited to, navigation through the structures, scaling and adjusting the visibility of individual components, applying cross-sectional views and simulating surgical operations with basic mesh manipulations. Our framework is still work in progress. We here present preliminary results.

As a first step in practical evaluation, the resulting VR visualizations were preliminary assessed using HMDs by two certified senior neurosurgeons. We demonstrate here the capabilities of our framework on data of one healthy subject and two patients suffering from intracranial tumors.

\section{Related Work}

\paragraph{Visualization and Segmentation}
With the rapid development of computer graphic capabilities, modeling software and structural neuroimaging techniques, some research groups successfully implemented realistic 3D visualizations of brain structures. Kin et al.~\cite{kin2009prediction} demonstrated the superiority of 3D visualization by fusing MRA and fast imaging employing steady-state acquisition data acquired from patients suffering from neurovascular compression syndrome. They implemented a 3D interactive VR environment, where neurosurgeons were efficiently able to detect the simulated offending vessels.
Furthermore, Chen et al.~\cite{chen2011human} fused fMRI activation maps and neuronal white matter fiber tractography acquired by magnetic resonance diffusion tensor imaging (DTI), and created precise VR visualizations that aim to describe cortical functional activities in addition to neuronal fiber connectivity. H{\"a}nel et al.~\cite{hanel2014interactive} effectively visualized the progression of brain tissue loss in neurodegenerative diseases in an interactive VR setting. Both transparency based and importance-driven volume rendering were used to highlight spatial localization and temporal progression. Abhari et al.~\cite{abhari2015visual} emphasized the importance of contour enhancement and stereopsis in visualizing vascular structures. Their study showed that volume rendering and a non-photorealistic shading of non-contrast-enhanced MRA based models increase the perception of continuity and depth. Recently, in a collaboration between anatomists, 3D computer graphics artists and neurosurgeons, Hendricks et al.~\cite{hendricks2018operative,hendricks2018cerebrovascular} exploited state of the art computer graphics and digital modeling technologies to create intricate, anatomically realistic and visually appealing virtual models of the cranial bones and cerebrovascular structures.

Furthermore, brain image segmentation benchmarks have been established to accelerate research and assessment of segmentation methods, including BRATS\cite{menze2015multimodal} for brain tumors, MRBrainS\cite{mendrik2015mrbrains} for targeting brain tissue segmentation and ISLES\cite{maier2017isles} for stroke lesions. In addition to traditional methods, recent advances in deep learning algorithms and artificial neural networks architectures have shown promising results in image segmentation tasks. Reviews regarding the implementation of deep learning in medical imaging and computer-aided diagnosis can be found in~\cite{litjens2017survey,shen2017deep}.

\paragraph{Commercial VR frameworks}
Several VR integrated hardware and software platforms have been established in recent years in the field of neurosurgical planning, medical training and education. The Dextroscope\textregistered presents brain components in a stereoscopic visualization as 3D holographic projections that can be manipulated in real-time to simulate and plan neurosurgeries. It has been used in a variety of intervention simulations including arteriovenous malformations resection~\cite{ng2009surgical,wong2009stereoscopic}, cerebral gliomas~\cite{qiu2010virtual}, cranial base surgery~\cite{kockro2009virtual,yang2009clinical} clipping of intracranial aneurysms~\cite{wong2007craniotomy} and meningioma resection~\cite{low2010augmented}. NeuroTouch, a training framework, incorporates haptic feedback designed for the acquisition and assessment of technical skills involved in craniotomy-based procedures and tumor resection~\cite{alotaibi2015assessing,alzhrani2015proficiency,azarnoush2015neurosurgical,delorme2012neurotouch}. ImmersiveTouch\textregistered also provides platforms in AR and immersive VR environments with haptic feedback for neurosurgical training~\cite{alaraj2013role}, mainly for cerebral aneurysm clipping~\cite{alaraj2015virtual} and ventriculostomy~\cite{schirmer2013virtual,luciano2005second}. Platforms that lack scientific assessment include NeuroVR\footnote{\url{https://caehealthcare.com/surgical-simulation/neurovr/}}, Surgical Theater\footnote{\url{https://www.surgicaltheater.net/}}, Osso VR\footnote{\url{http://ossovr.com/}} and Touch Surgery\footnote{\url{https://www.touchsurgery.com/}}. More comprehensive reviews regarding the development and utilization of VR environments for neurosurgical purposes can be found in~\cite{robison2011man,schirmer2013evolving,chan2013virtual,pelargos2017utilizing}. The main differentiation of our work in relation to the above is that we focus on the use of off-the-shelf consumer hardware and open-source software. These two criteria ensure that our work can potentially be used by anyone, for non-commercial purposes. Finally, quantitative scientific assessment of our framework is planned upon achievement of the first RC.

\section{Methods}

\subsection{Data}
We have used the MRI and CT data of one healthy subject (S1, MRI only) and 2 patients (P1-2) as our pilot cohort to construct detailed 3D models to be viewed and interacted with in VR. Table~\ref{tab:overviewSubjects} gives an overview of the cohort. The MRI and CT data we used for the segmentation is listed in table~\ref{tab:overviewDataSets}. These data were acquired as part of the routine pre-surgical diagnostic procedure. The selection of imaging sequences is purely driven by the diagnostic procedure, which results in different sets of data per patient.

\begin{table*}[h!]
  \caption{Overview of pilot cohort}
  \label{tab:overviewSubjects}
  \scriptsize%
	\centering%
  \begin{tabu}{%
	r%
	*{3}{c}%
	}
  \toprule
   & Age & Sex &  Diagnosis  \\
  \midrule
    Subject 1 \cite{fiederer2016role}& 27 & M & no history of neuro-psychiatric disease\\
    Patient 1 & 23 & F & pilocystic astrozytoma, temporo-mesial, left\\
    Patient 2 & 70 & F & undefined tumor with cystic components, pontomedullary junction, right \\

  \bottomrule
  \end{tabu}%
  
\end{table*}

\begin{table*}[h!]
  \caption{Overview of imaging data 
  }
  \label{tab:overviewDataSets}
  \scriptsize%
	\centering%
  \begin{tabu}{%
	r%
	*{3}{c}%
	}
  \toprule
  Imaging modality & Subject 1\cite{fiederer2016role} &  Patient 1 &   Patient 2 \\
  \midrule
    \textbf{Whole-head}\\
    CE-MRI T1 MPRAGE (T1 MRA) & 
    - &
    0.5$\times$0.5$\times$1\,mm & 0.5$\times$0.5$\times$1\,mm \\
    
    CE-MRI T2 SPACE (T2 MRA) &
    - &
    1$\times$1$\times$1\,mm & 0.5$\times$0.5$\times$1\,mm \\
    
    MRI T1 MPRAGE &
    0.6$\times$0.6$\times$0.6\,mm &
    1$\times$1$\times$1\,mm & 0.488$\times$0.488$\times$1\,mm \\
    
    MRI T2 SPACE &
    0.6$\times$0.6$\times$0.6\,mm &
    1$\times$1$\times$1\,mm & - \\
    
    MRI FLAIR &
    - &
    1$\times$1$\times$1\,mm & 1.016$\times$1.016$\times$1\,mm \\
    
    CT &
    - &
    0.5$\times$0.5$\times$0.5\,mm &
    - \\
    \\
    \textbf{Region of Interest} \\
    Skull-base MRI TOF (TOF MRA) &
    - &
    - & 0.352$\times$0.352$\times$0.7\,mm \\
    
    Skull-base CT &
    - &
    - & 0.396$\times$0.396$\times$0.400\,mm \\
    
    CT angiography (CTA) &
    - &
    0.25$\times$0.25$\times$0.25\,mm & -  \\

  \bottomrule
  \end{tabu}%
  \\
  CE: contrast-enhanced (ProHance\textregistered), MPRAGE: magnetization prepared rapid gradient echo, MRA: magnetic resonance angiography, SPACE: sampling perfection with application optimized contrasts using different flip angle evolution, 
  FLAIR: fluid-attenuated inversion recovery, TOF: time of flight
  
\end{table*}

\subsection{Segmentation}

\begin{figure*}[h!]
    \centering
    \includegraphics[width=0.9\linewidth]{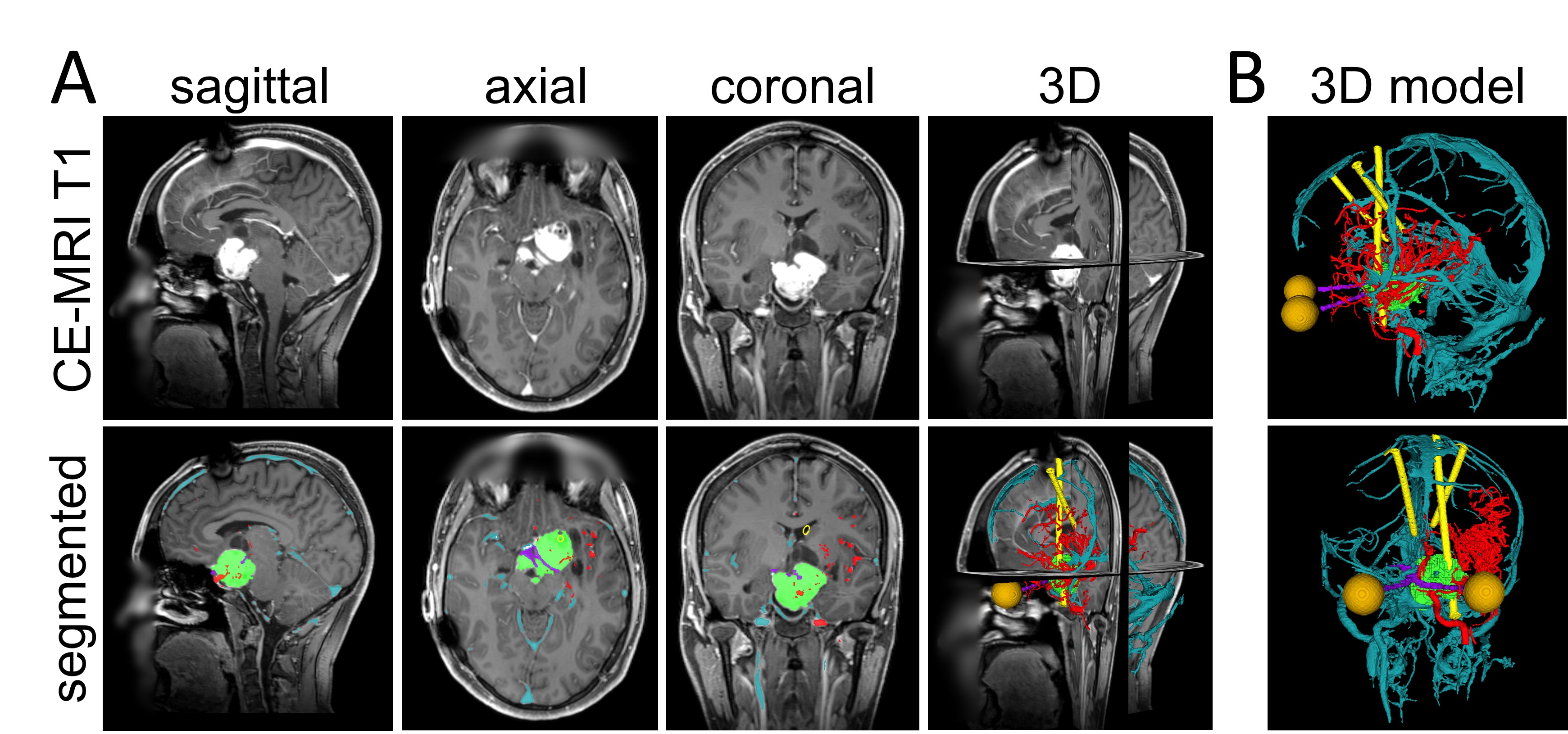}
    \caption{Example of segmentation in Slicer with data of P1 (face area blurred).
    A) Contrast-enhanced T1 MRI data (top row) and segmentation overlay (bottom row), including vessels segmented from MRA (cyan) and CTA (red), catheters (yellow), tumor (green), eyes (orange), and optic tract (purple). B) Resulting 3D model in Slicer.
    }
    \label{fig:Slicer}
\end{figure*}

\begin{figure*}[h!]
    \centering
    \includegraphics[width=\linewidth]{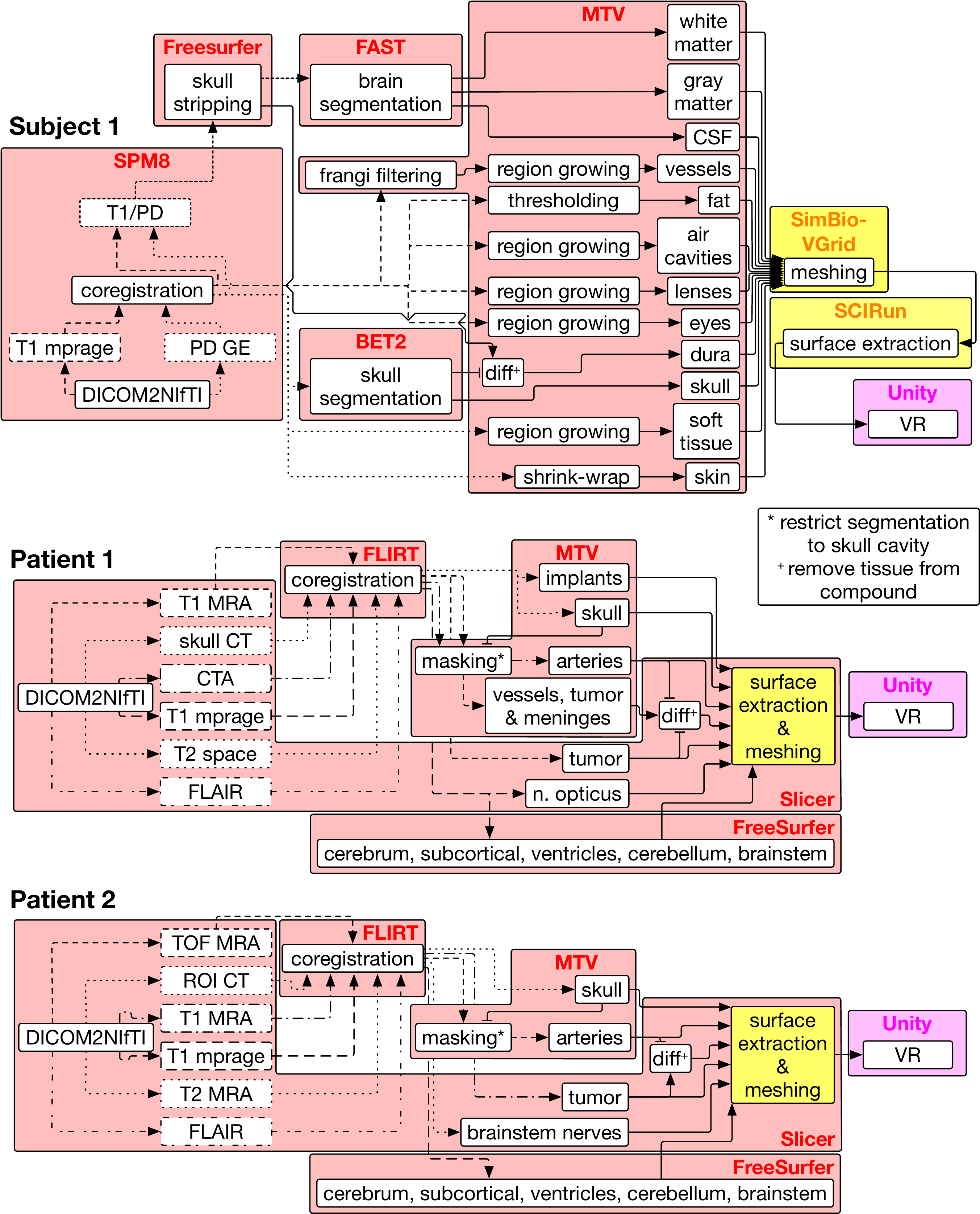}
    \caption{Segmentation pipelines. Colours indicate the components of the overarching framework, as used in Fig.~\ref{fig:Framework}. Boxes enclose the steps performed with individual open-source frameworks. The texture of the lines indicate which imaging modality was used in which step. MTV (MRI TV) is our own, Matlab-based MRI processing framework.
    }
    \label{fig:seg_pipelines}
\end{figure*}

To produce high-quality results on the clinical routine data, we adapt the segmentation pipeline used to create the model of S1\cite{fiederer2016role} for each patient individually. The pipeline, which is still work in progress, had to be adapted on an individual basis to account for the wide range of available data, their variable quality and the different pathologies. As our, yet unpublished, deep-learning-based segmentation pipeline reaches maturity we plan to replace the current pipeline to have a single, fully automated, pipeline covering all segmentation cases. First, we converted the imaging data from the DICOM to the NIfTI format. Then, we co-registered and resliced every volume to the T1 MRA (P1) or the TOF MRA (P2) using FSL FLIRT~\cite{greve_accurate_2009, jenkinson_improved_2002, jenkinson_global_2001} with default parameters. We segmented the skull, intracranial catheters and implants (P1), from the CT volumes using our own semi-automatic multi-stage region growing algorithm. We mask the CTA \& MRA volumes with the dilated skull to prevent segmentation spillage during the semi-automatic multi-stage region growing segmentation of blood vessels (P1: arteries from CTA, mixed vessels from T1 MRA. P2: arteries from TOF MRA)
. FreeSurfer\footnote{Documented and freely available for download online (\url{http://surfer.nmr.mgh.harvard.edu/})} was used to create the segmentation of the cerebral cortex~\cite{dale1999cortical}, sub-cortical structures, cerebellum, ventricles and brainstem~\cite{fischl2004automatically}. Tumors, brainstem nerves (P2), optic tract (P1) and eyes (P1) were segmented manually using T1 MRA, T2 MRA and T1, respectively, imported into 3D Slicer~\cite{kikinis_3d_2014, fedorov_3d_2012} (Fig. \ref{fig:Slicer}). Fig.~\ref{fig:seg_pipelines} gives a graphical representation of the original segmentation pipeline used for S1 and the individually adapted segmentation pipelines of P1 and P2. When running on an i7-8700K CPU (12 3.7\,GHz threads), the semi-automated part of the pipeline (without manual segmentation) generally takes 6-7\,h to complete, with FreeSurfer taking 5-6\,h.
\subsection{Visualization}
We used Blender\footnote{\url{https://www.blender.org/}} to calculate inner and outer surface normals to enable inner and outer visualization of the volumetric models. Also, smooth shading was applied to modify shading calculations across neighbouring surfaces to give the perception of smooth surfaces without modifying the model's geometry.

We developed our VR 3D environment using the Unity3D game engine. Unity3D is equipped with easy-to-use navigation and interaction tools designated to maximise the user experience. We designed the environment as follows: 2 directional real-time light sources and environment lighting ambient mode set to real-time global illumination were used to enhance illumination from all directions and create realistic reflections with various model orientations in real-time. The standard Unity3D shader was used. Perspective projection rendering was used to improve the sense of depth. Additionally, forward rendering was used for cost-effective calculations and applying transparency shaders, which will be implemented in future versions. Another advantage of using forward rendering in VR is the ability to use multisample anti-aliasing as a pre-processing step, which is a very useful technique to improve surface smoothness and minimize jagged edges.
The reflectivity and light responses of the surfaces of each layer was set to give a glossy plastic texture by maximizing smoothness, and minimizing the metallic property, which also eliminated reflections when looking at the surface face-on. An Oculus Rift\footnote{\url{https://www.oculus.com/rift/}} headset and two Touch controllers bundle were used for a fully immersive VR environment.

\subsection{User interface}
As seen in Fig.~\ref{fig:modelInteraction_Patient2}A, user interface (UI) elements holding the layers' names and control modes are generated inside a canvas that can be switched on/off. The canvas is rendered in a non-diegetic fashion, i.e., heads up display, to enable the canvas to track and stay within the field of vision in the world space with fixed distance. This would improve eye focus on the UI elements.

Two modes of interaction were implemented effectively, the raycaster interaction (RI) mode and the controllers tracking interaction (CTI) mode. In RI mode, UI elements in the canvas include both layers' names and control modes, namely, Move Avatar, which enables translational movement in three perpendicular axes, i.e., surge, heave and sway, relative to the gaze forward direction. The second control mode is Rotate Model, which allows yaw and pitch rotations. The Scale Model UI element enables the user to scale the model up and down. Finally, a Reset option is provided to return the model and the avatar to their initial state. Furthermore, a graphic raycaster is used to interact with the UI elements through unity's event system. 

The CTI mode is based on the distance and the relative angles between the controllers, allowing yaw, pitch, roll rotations and model scaling, by creating circular and linear motions with the controllers, respectively. The canvas in this mode only includes the layers' names. CTI interactions have been suggested to us by the experienced neurosurgeons having preliminarly evaluated the framework.

Lastly, a cross-sectional view of the model is provided by controlling three planes.
In that way, surgeons can compare cross-sections of the segmented models with the corresponding cross-sections in the raw MRI or CT data.

\section{Results}

Videos showcasing our results are available on our website \url{https://www.tnt.uni-freiburg.de/research/medical-vr-segmentation/aveh_2019}.

\subsection{Subject 1}
\begin{figure}[h!]
\centering{\includegraphics[width=0.47\textwidth]{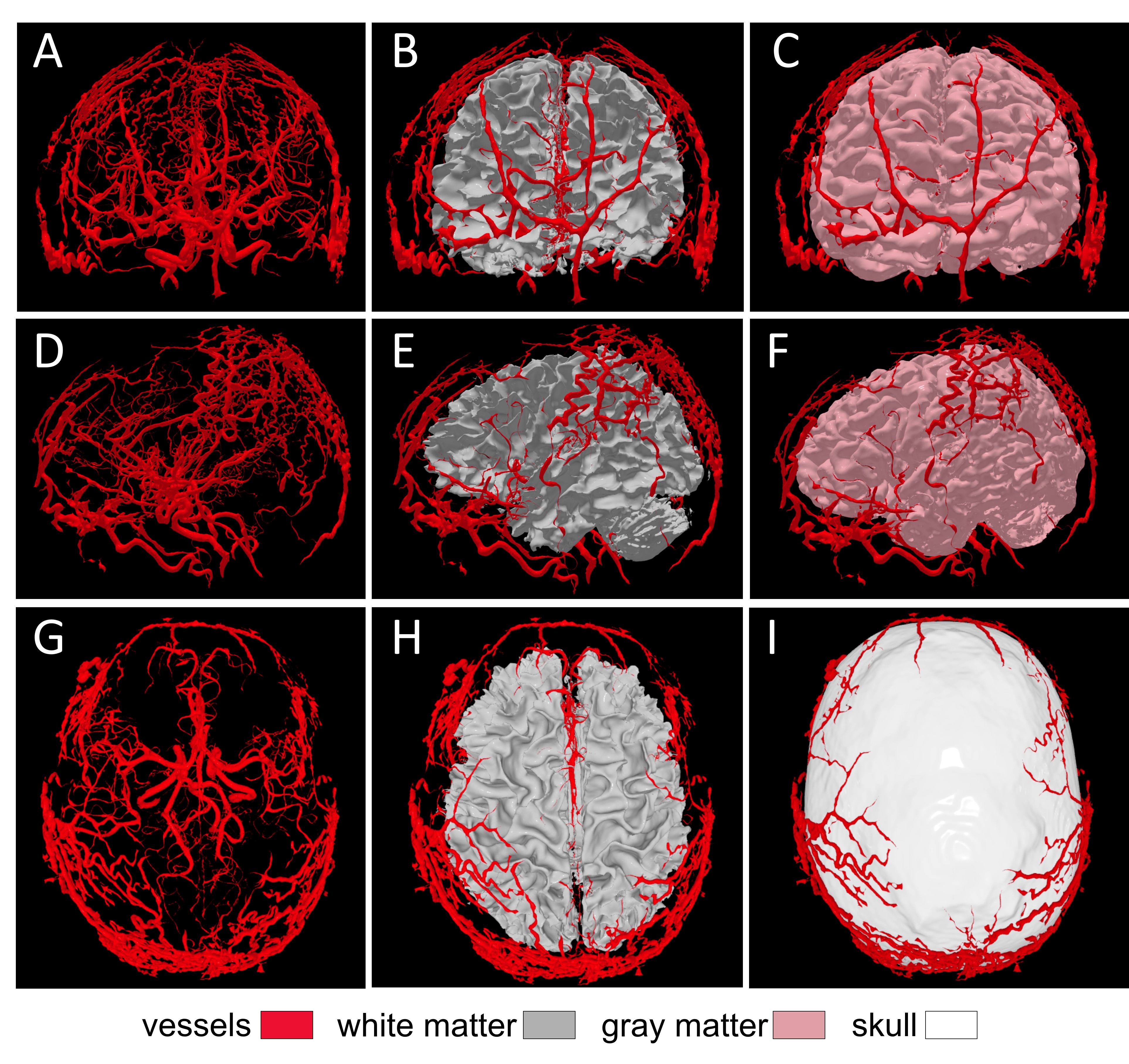}}
\caption[Model overview S1]
{Model overview of S1.
The upper row shows a frontal view of the cerebrovascular tree (A) with white matter (B) or gray matter (C). The middle row displays a view from the left side of the vessels (D) with white (E) or gray matter (F). The lower row includes views from the top of the vessels (G) with white matter (H) or skull (I).}
\label{fig:modelOverview_Subject1}
\end{figure}

From the data of S1, the following structures were segmented and modeled:
Cerebrovascular tree, skull, CSF, dura mater, eyes, fat, gray matter, white matter, internal air, skin, and soft tissue. All models are compatible with further extension by adding more segmented areas or structures. Fig.~\ref{fig:modelOverview_Subject1} shows an overview of S1's model in Unity.
\subsection{Patient 1}
\begin{figure}[h!]
\centering{\includegraphics[width=0.47\textwidth]{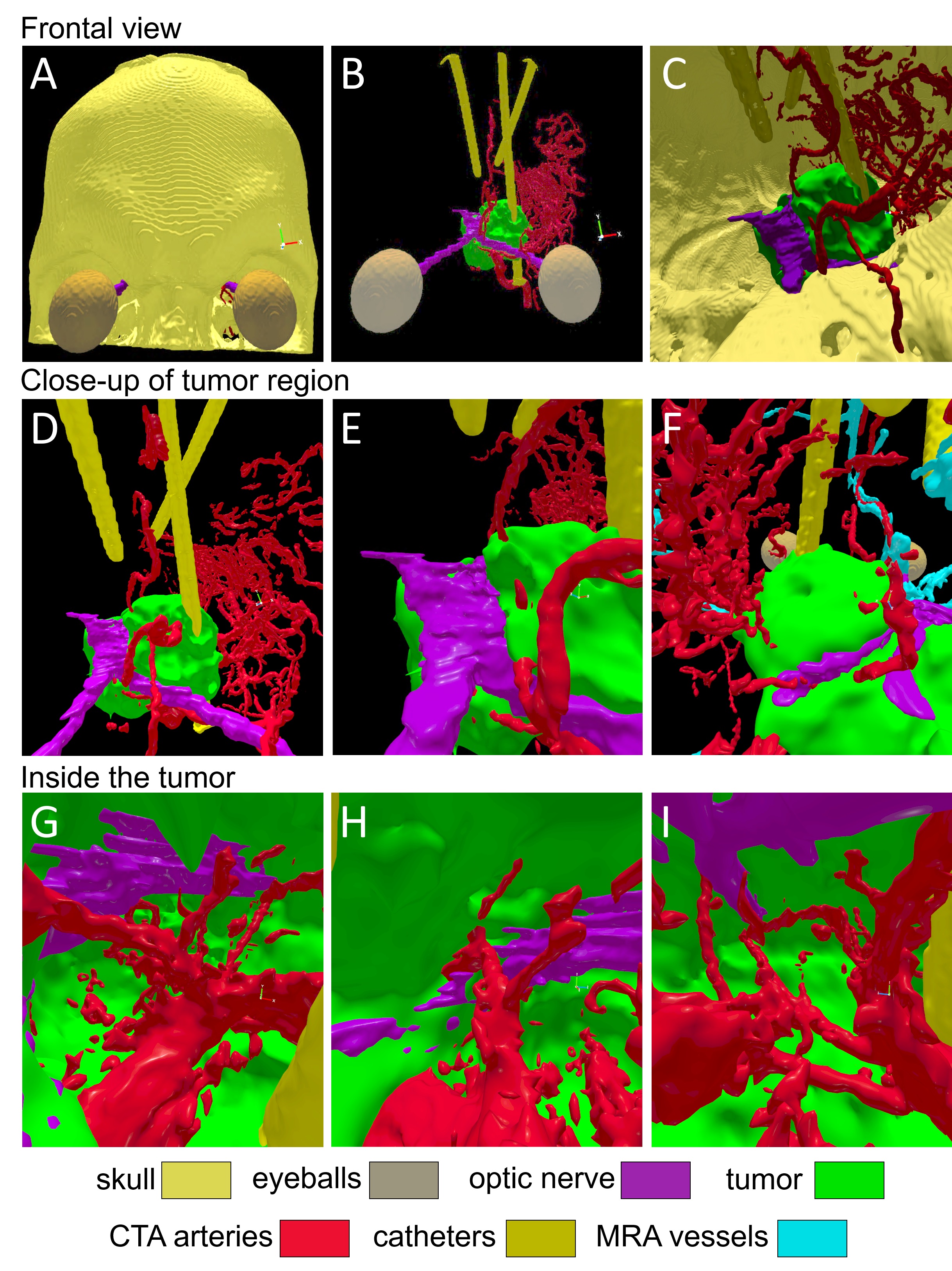}}
\caption[Model overview of patient 1]
{Model overview of P1.
The upper row contains screenshots showing a frontal view A) with skull, B) without skull, and C) inside the skull cavity.
The middle row contains close-up pictures of the tumor region as D) frontal view showing the tumor’s growth displacing and enveloping optic chiasm, E) frontal view showing tumor’s growth enveloping arteries, and F) posterior view showing tumor’s growth enveloping arteries and optic tract.
The lower row contains screenshots from within the tumor. G) Frontal view centered on C1/A1/M1 junction, H) lateral view centered on C1/A1/M1 junction, and I) lateral view centered on C1/A1/M1 junction. Please note that the model of P1 was not smoothed to asses whether neurosurgeons prefer to see the models in their CT/MRI ground truth form or in a smoothed form.
}
\label{fig:modelOverview_Patient1}
\end{figure}

From the data of P1, the following anatomical structures were segmented and modeled, as they were of importance to the tumor region:
Meninges, catheters, optic tract, MRA vessels, CTA arteries, skull, tumor, eyes, cerebral and cerebellar gray and white matters, subcortical structures, ventricles and brainstem. Not all structures were included into the VR representation. Fig.~\ref{fig:modelOverview_Patient1} shows an overview of the model created from data of P1.

\subsection{Patient 2}
\begin{figure}[h!]
\centering{\includegraphics[width=0.45\textwidth]{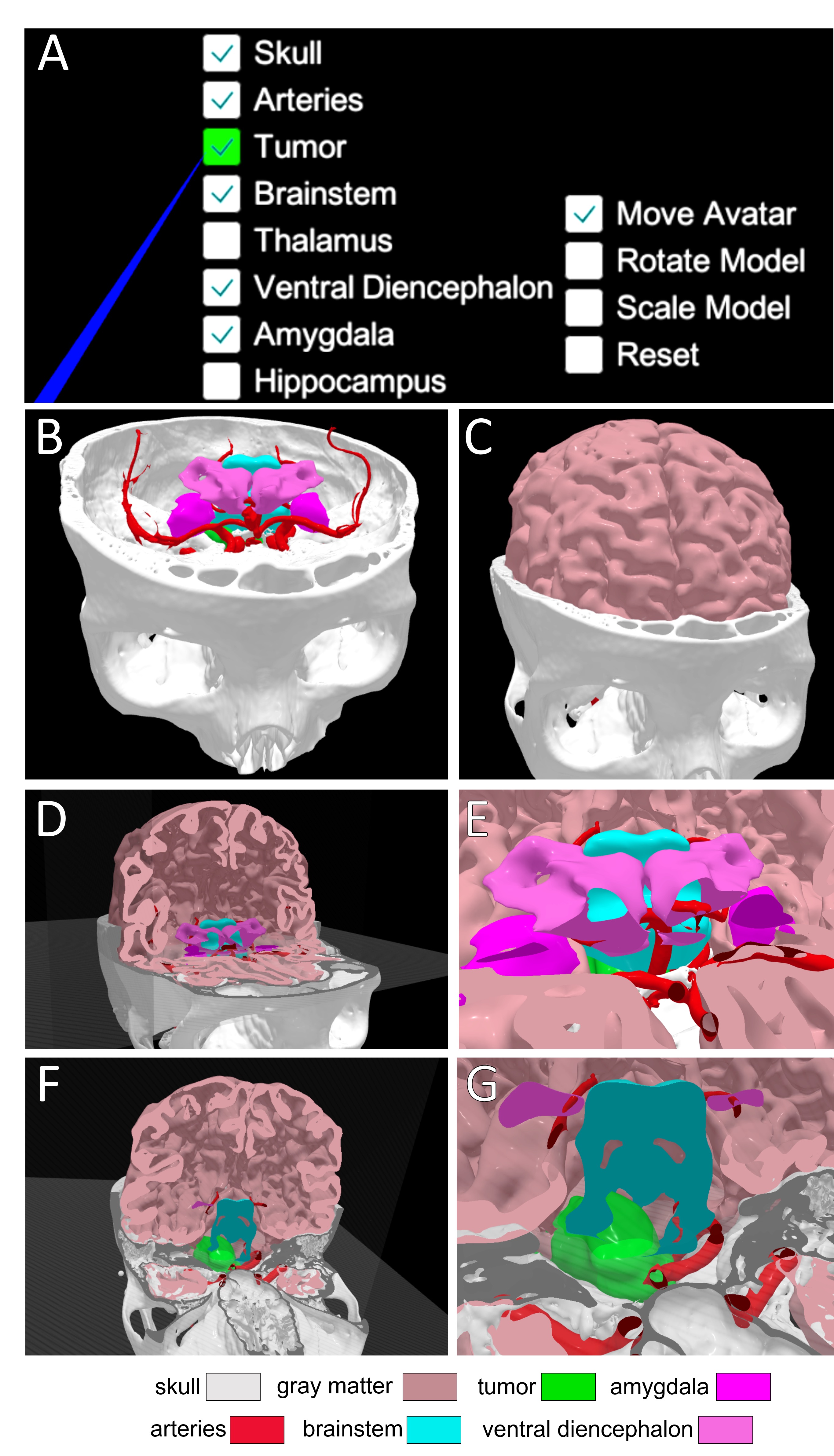}}
\caption[Interaction example patient 2]
{Interaction example in data of P2.
A) With the right Touch controller, the user can choose which anatomical structures they want to display and if they want to move, rotate or scale the model (excerpt of menu).
Examples of the displayed anatomical structures B) without gray matter and C) with gray matter.
Further, the user can apply and slice cross-sectional views in horizontal and vertical plane using the buttons and joystick on the left Touch controller. Pictures D-E show examples of different cross-sectional views of the whole head (D, F) or after zooming in (E, G).}
\label{fig:modelInteraction_Patient2}
\end{figure}

From the data of P2, the following anatomical structures were segmented and modeled:
Intracranial arteries, skull, tumor, cerebral and cerebellar gray and white matters, subcortical structures, ventricles and brainstem and brainstem nerves. Not all structures were included into the VR representation. Fig.~\ref{fig:modelInteraction_Patient2} shows an examplary interaction with the model of P2, including application of cross-sectional views.

\subsection{First Evaluation}
\begin{figure*}[h!]
\centering{\includegraphics[width=\textwidth]{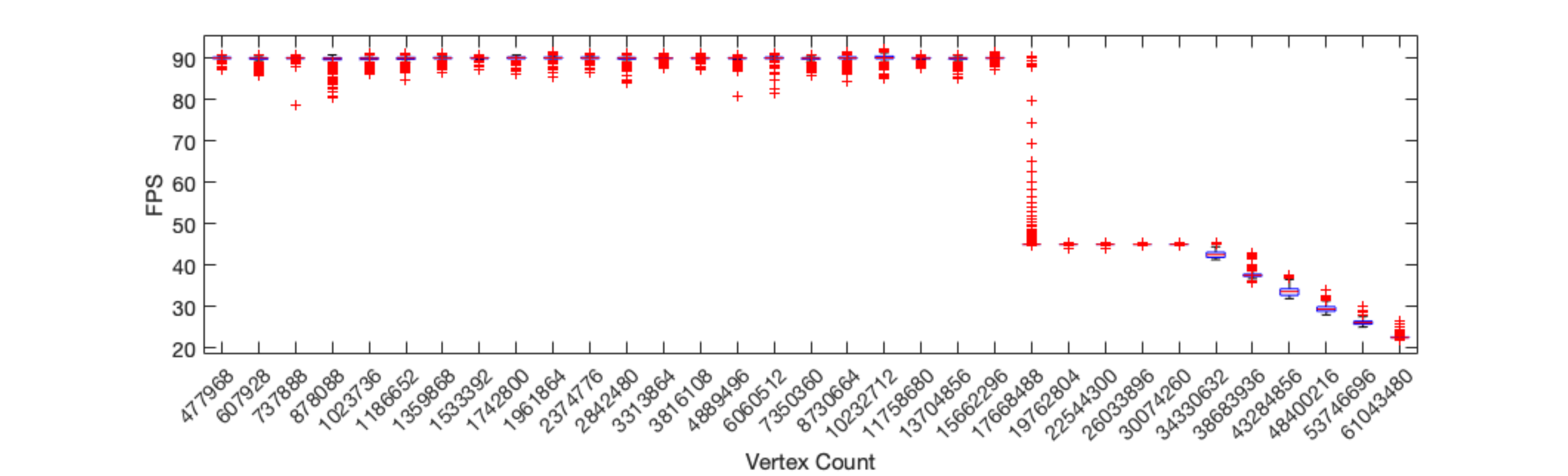}}
\caption[Preliminary Performance Evaluation of the VR Visualization]
{Preliminary Performance Evaluation of the VR Visualization. Performance is stable across interactions but decreases with increasing number of vertices passed to the GPU. Vertices of multiple models are accumulated to showcase the performance. Occlusion culling of vertices not visible to the user still has to be implemented to further increase the performance.}
\label{fig:GPU}
\end{figure*}

\paragraph{Performance}We preliminarly evaluated the performance of the VR visualizations by recording the frame delta time and the count of vertices passed to the GPU for different model configurations and interactions. A NVIDIA GTX 1080 Ti is used as GPU. Frame rates never dropped below 30 FPS for any given model, which ensures that the interaction stays smooth under any load. Results are visualized in Fig.~\ref{fig:GPU}.

\paragraph{Physicians}Two experienced neurosurgeons (senior physicians) inspected the 3D models postoperatively in VR prior to inspecting the 2D images in 2D. They reported being surprised by the accuracy of the models, the degree of immersion and the speed at which they were able to assimilate the information. During the subsequent inspection of the 2D images, the neurosurgeons were able to quickly verify their VR-based internal 3D representation and successfully transferred these insights from VR to 2D.

\section{Discussion}


In the present study we have designed and performed a first preliminary evaluation of, to the best of our knowledge, the first dedicated, research software immersive VR framework for neurosurgery.This also shows that existing open-source tools and consumer hardware can be used to create a research framework for VR neurosurgery.

The requirements in the context of neurosurgery are highly specialized. For example, the exact spatial interrelations of blood vessels, pathological structures such as a brain tumor, and both gray and white matter brain structures can be decisive for planning the detailed neurosurgical approach, such as the extent of a planned resection. In parallel to ongoing development by commercial solutions, we believe that there is also a need for academic research into topics including optimized segmentation algorithms, intuitive visualization, and interface design. We anticipate that the availability of frameworks based on open-source tools will facilitate future research and development into VR applications in the context of neurosurgical planning, as well as teaching and training. 

We have showcased the working procedure of our framework on data of one healthy subject and two tumor patients. Due to different requirements in regard to the final models, the processing pipeline had to be personalised and differed between the cases. For automated segmentation and modeling methods, this need for personalisation represents a big challenge. In order to solve this issue, we plan to apply state-of-the-art machine learning techniques, e.g., deep convolution neural networks, in the future.

To further improve our framework, we will devise a more intuitive, user-friendly and interactive UI, and implement mesh manipulations to mimic neurosurgical procedures, e.g., sawing and burring as demonstrated by Arikatla~\etal \cite{arikatla2018high}. Moreover, transparent shaders will be implemented in the near future to improve spatial perception of overlaying brain structures.

\section{Conclusion}

Here we present first evidence that 3D models of pre-neurosurgical data presented in VR could provide experienced neurosurgeons with additional information and could potentially ease the preparation of surgery.
Thus VR is a promising tool to speed up and improve the preparation of surgery, reduce time needed for surgery and in doing so improve the outcomes.
To assess the true potential of VR surgical planning, in the context of improving surgical outcomes, we are currently preparing a quantitative evaluation of our framework and are planning a retrospective study of poor outcome patients.

\acknowledgments{
This work was supported by the German Research Foundation grant EXC 1086.
}

\bibliographystyle{abbrv-doi}

\bibliography{bibliography}
\end{document}